\documentclass[11pt]{article}
\usepackage{cospar}
\usepackage{url}

\usepackage{graphicx}
\usepackage[figuresright]{rotating}

\title{FORMATION AND MIGRATION OF TRANS-NEPTUNIAN OBJECTS AND ASTEROIDS}

\author{S.I. Ipatov\address{(1) Institute of Applied Mathematics, 
Miusskaya sq. 4, Moscow 125047, Russia; E-mail: siipatov@hotmail.com; 
(2) NRC/NAS senior research associate, NASA/GSFC, Mail Code 685, 
Greenbelt, MD 20771, USA (current address)}}

\begin{document}

\maketitle

\begin{abstract}

The evolution of thousands of orbits of Jupiter-family comets and asteroids under the gravitational influence of planets was calculated. Comparison of the results obtained by a symplectic method with those obtained by direct integration showed that a symplectic method is not always good for investigations of the orbital evolution of such bodies. Basing on the results of orbital evolution of bodies, we concluded that a considerable portion of near-Earth objects could have come from the trans-Neptunian region. Some large trans-Neptunian objects could be formed by the compression of rarefied dust condensations, but not by the accumulation of smaller planetesimals.
\end{abstract}

\section*{1. FORMATION OF TRANS-NEPTUNIAN OBJECTS AND ASTEROIDS}

It is considered by many authors that a dust disk around the forming Sun became thinner until its density reached a critical value about equal to the Roche density. At this density, the disk became unstable to perturbations by its own self-gravity and developed dust condensations. These initial condensations coagulated under collisions and formed larger condensations, which compressed and formed solid planetesimals. Usually it is considered that asteroids and trans-Neptunian objects (TNOs) were formed by accumulation of smaller planetesimals. As it was obtained by several authors, the process of accumulation of TNOs from smaller planetesimals needs small ($\sim$0.001) eccentricities and a massive belt.

     Our runs showed that maximal eccentricities of typical TNOs always exceed 0.05 during 20 Myr under the gravitational influence of the giant planets. Gas drag could decrease eccentricities of planetesimals, and the gravitational influence of the forming giant planets could be less than that of the present planets. Nevertheless, to our opinion, it is probable that, due to the gravitational influence of the forming giant planets and migrating planetesimals, small eccentricities of TNOs could not exist during all the time needed for the accumulation of TNOs with diameter $d>100$ km.

     Therefore, we support the Eneev's suggestion that TNOs with $d\ge100$ km moving now in not very eccentric orbits could be formed directly by the compression of large rarefied dust condensations (with $a>30$ AU), but not by the accretion of smaller solid planetesimals. The role of turbulence could decrease with an increase of distance from the Sun, so, probably, condensations could be formed at least beyond Saturn's orbit.

     Probably, some planetesimals with $d\sim100-1000$ km in the feeding zone of the giant planets and even large main-belt asteroids also could be formed directly by the compression of rarefied dust condensations. Some smaller objects (TNOs, planetesimals, asteroids) could be debris of larger objects, and other such objects could be formed directly by compression of condensations. Even if at some instants of time at approximately the same distance from the Sun, the dimensions of initial condensations, which had been formed from the dust layer due to gravitational instability, had been almost identical, there was a distribution in masses of final condensations. As in the case of accumulation of planetesimals, there could be a "run-away" accretion of condensations. 

     A small portion of planetesimals from the feeding zone of the giant planets that entered into the trans-Neptunian region could be left in eccentrical orbits beyond Neptune and became so called ''scattered disk objects'' (SDOs). The end of the bombardment of the terrestrial planets could be caused mainly by those planetesimals that had become SDOs. 

     Our estimates [1-2] showed that collisional lifetimes of TNOs and asteroids are about 1 Gyr. Typical TNOs can be even more often destroyed by SDOs than by other TNOs. Less than 1\% of 100-km TNOs and most of 1-km TNOs could be destroyed during last 4 Gyr. Collisional lifetimes are smaller for smaller objects. The collisional lifetime of 1-m asteroid is about 1 Myr. 

\section*{2. ORBITAL EVOLUTION OF JUPITER-FAMILY COMETS AND RESONANT ASTEROIDS}

The motion of TNOs to Jupiter's orbit was investigated by several authors [3]. In [2] the quantative analysis of migration of TNOs into the near-Earth space, including probabilities of their collision with the Earth, was based on the results of the evolution of 48 orbits of Jupiter-crossing objects (JCOs). In the present paper we consider the orbital evolution of thousands JCOs. 

\begin{table}

\caption{Values of $T$ and $T_J$ (in Kyr), $P_r$, and $r$ obtained by 
the BULSTO code (Venus=V, Earth=E, Mars=M)}

$ \begin{array}{lllllcccccccc} 

\hline	

  & & &&&$V$ & $V$ & $E$ & $E$ & $M$ & $M$ & &\\

\cline{6-11}

 & N& a& e&i&P_r & T & P_r & T & P_r & T & r & T_J \\

\hline

n1&1900& &&& 2.42 & 4.23 & 4.51 & 7.94 &  6.15 & 30.0 & 0.7 & 119\\
$2P$ & 501 &2.22 &0.85&12&226 &504  & 162 & 548 &  69.4  & 579 & 19. & 173\\
$9P$ & 800&3.12 &0.52&10&1.34 & 1.76 & 3.72 & 4.11 &     0.71 & 9.73 & 1.2 &96\\
$10P$ & 2149&3.10&0.53&12&28.3 & 41.3 & 35.6 & 71.0 &   10.3 & 169. & 1.6 &122\\
$22P$ & 1000&3.47 &0.54&4.7&1.44&2.98&1.76 & 4.87 &     0.74 & 11.0 & 1.6 &116\\
$28P$ & 750 &6.91&0.78&14& 1.7 & 21.8 & 1.9 & 34.7 & 0.44 & 68.9 & 1.9 &443\\
$39P$ & 750 &7.25&0.25&1.9& 1.06&1.72&  1.19& 3.03 & 0.31 & 6.82 & 1.6 &94\\
$total$& 7850&&&& 17.9 & 37.7 & 18.8 & 51.5 &  5.29 & 85.9 & 2.6 &116\\
3:1& 288 & 2.5&0.15&10& 940 &1143& 1223& 1886 &       371& 3053 & 2.3 &227\\
5:2& 288 & 2.82&0.15&10& 95.8 &170 & 160 & 304 &       53.7 & 780 & 1.0 &232\\

\hline
\end{array} $ 
\end{table}

     Below we present the results of the evolution of bodies under the gravitational influence of planets (except for Mercury and Pluto) during a time span $T_S\ge10$ Myr. We used the integration package from [4]. In the present section (including Table 1 and Figs. 1-2) we present the results obtained by the Bulirsh-Stoer method (BULSTO code), and in the next section we discuss the difference between the results obtained by BULSTO and a symplectic method (RMVS3 code). For BULSTO the error per integration step was taken (depending on the run) to be less than $\varepsilon=10^{-9}$ or $\varepsilon=10^{-8}$, or some value between these two values.

     In the first series of runs (denoted as $n1$) we considered by BULSTO the orbital evolution of $N=1900$ JCOs moving in initial orbits close to those of 20 real JCOs with period 
$5<P_a<9$ yr. In other series of runs we considered initial orbits close to those of one comet (2P, 9P, 10P, 22P, 28P, or 39P). For 2P runs we included Mercury. We also investigated the 
evolution of asteroids initially moving in the resonances 3:1 and 5:2 with Jupiter. For JCOs we varied only the initial mean anomaly, and for asteroids we varied also initial value of the 
longitude of the ascending node. The approximate values of initial orbital elements ($a$ in AU, $i$ in deg) are presented 
  in Table 1.











     In our runs planets were considered as material points, but, based on orbital elements obtained with a step 500 yr, we calculated the mean probability $P=P_\Sigma/N$ ($P_\Sigma$ is the probability for all $N$ considered objects) of a collision of an object with a planet and the mean time $T=T_\Sigma/N$ during which perihelion distance $q$ of an object was less than a semi-major axis of the planet, and the mean time $T_J$ during which an object moved in Jupiter-crossing orbits. The values of $P_r=10^6P$, $T_J$ and $T$ are shown in Table 1. Here $r$ is the ratio of the total time interval when orbits are of Apollo type ($a>1$ AU, $q=a(1-e)<1.017$ AU) at $e<0.999$ to that of Amor type ($1.017<q<1.33$ AU).

     For the calculation of $P$ in addition to the algorithm presented in [5], we take into account that a body moves in its orbit with a variable velocity at different distances r from the Sun, and it increases (compared to the approximation of a constant velocity) the time elapsed till a close encounter by a factor of $(2a/r-1)^{1/2}$. 

     The obtained results showed that the main portion of the probability of collisions of former JCOs with the terrestrial planets are due to a small ($\sim$0.1-1\%) portion of objects that moved during several Myrs in orbits with aphelion distances $Q<4.7$ AU. Some of them had typical asteroidal and near-Earth object orbits and could get $Q<3$ AU for millions of years. The ratio of the mean probability of a JCO with $a>1$ AU with a planet to the mass of the planet was greater for Mars than that for Earth by a factor of several (large $P$ for Mars at $n1$ was caused by one object with life-time equal to 26 Myr). The time (in Myr) spent by all 7852 JCOs and asteroids in several types of orbits is presented below:


$ \begin{array}{lllllllll} 


  & N& $IEOs$& $Aten$&  $Al2$& $Apollo$ & $Amor$ & a>5 $ AU$ \\


$JCOs$ & 7852 & 10 & 86 & 411 & 659 & 171 & 7100 \\

3:1 & 288   & 13  & 4.5 & 190 & 540 & 230 & 80 \\

5:2 & 288 & 0 & 0 & 2 & 90 & 90 & 230


\end{array} $


Here we consider inner-Earth objects (IEOs, $Q<0.983$ AU), Aten ($a<1$ AU and $Q>0.983$ AU), Al2 ($q<1.017$ AU and $1<a<2$ AU), Apollo, and Amor objects. The above times for Earth-crossing objects were mainly due to only several tens of objects. These orbits were on average more eccentric than those for actual objects. One former JCO, which initial orbit was close to that of 10P, got Aten orbits during 3.5 Myr (Fig. 1a-b), but the probability of its collision with the Earth from such orbits (0.3) was greater than that for 7850 considered former JCOs (0.15). It also moved during about 10 Myr (before its collision with Venus) in inner-Earth orbits and during this time the probability $P_V$ of its collision with Venus (0.7) was greater ($P_V\approx3$ for the time interval presented in Fig. 1a-b) than that for 7850 JCOs (0.14). Another object (Fig. 1c-d) moved in highly eccentric Aten orbits for 83 Myr, and its life-time before collision with the Sun was 352 Myr. Its probability of collisions with Earth, Venus and Mars was 0.172, 0.224, and 0.065, respectively. These two objects were not included in Table 1. The mean time $T_E$ during which a JCO was moving in Earth-crossing orbits equals $9.5\cdot10^4$ yr for 7852 considered JCOs, but it is less ($\approx8\cdot10^3$ yr) for $n1$. In [6] for $N=48$, $T_E=5\cdot10^3$ yr. Earlier several scientists obtained smaller values of $P$ than those presented in Table 1. The ratio $P_S$ of the number of objects collided with the Sun to the total number of escaped (collided or ejected) objects was:


$ \begin{array}{lllllll} 


0.0005 & 0.349 & 0 & 0.014 & 0.002 & 0.007 & 0 \\

n1 & 2P & 9P & 10P & 22P & 28P & 39P

\end{array}$


     Six and nine JCOs with initial orbits close to those of 10P and 2P, respectively, moved in Al2 orbits (with $1<a<2$ AU and $q<1$ AU) during at least 0.5 Myr each (five of them moved in such orbits during more than 5 Myr each). The contribution of all other objects to Al2 orbits was smaller. Some considered former JCOs spent a lot of time in the 3:1 resonance and with $2<a<2.6$ AU. Other objects got other Mars-crossing orbits for a long time. So JCOs can supply bodies to the regions which considered by many scientists [7] to belong to the main sources of near-Earth objects (NEOs). 

\begin{figure}
\includegraphics[width=91mm]{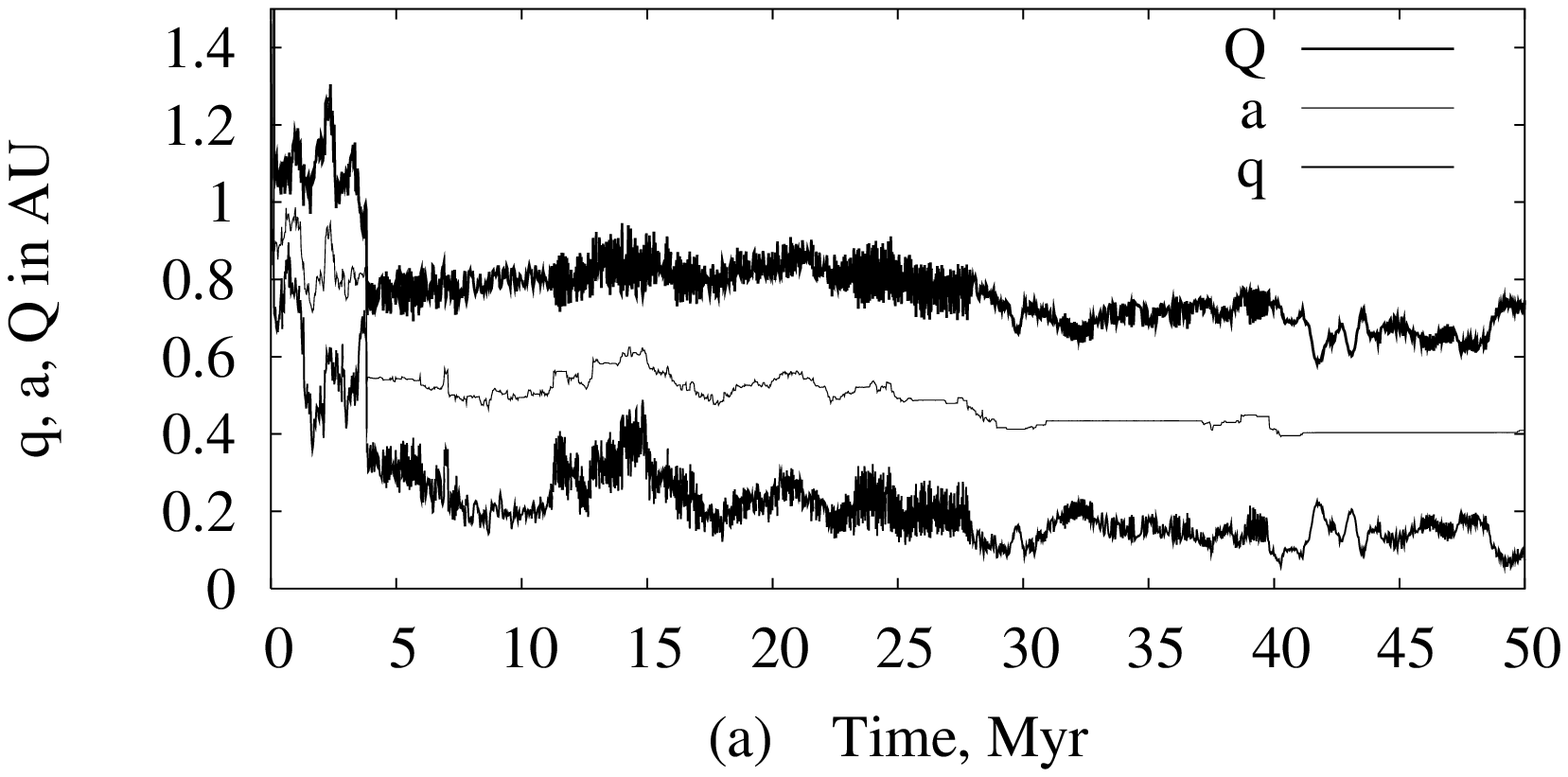}
\includegraphics[width=91mm]{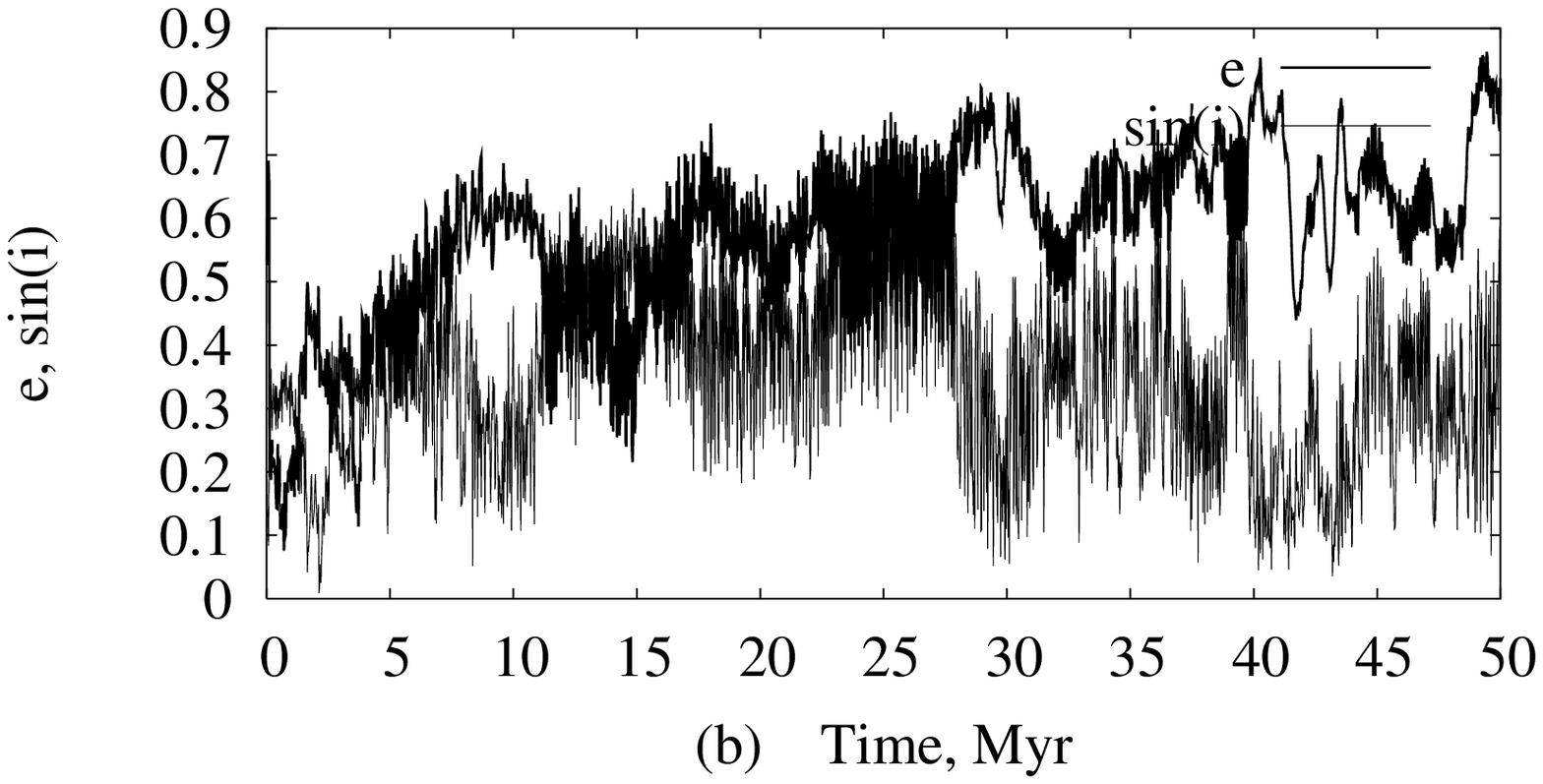}
\includegraphics[width=91mm]{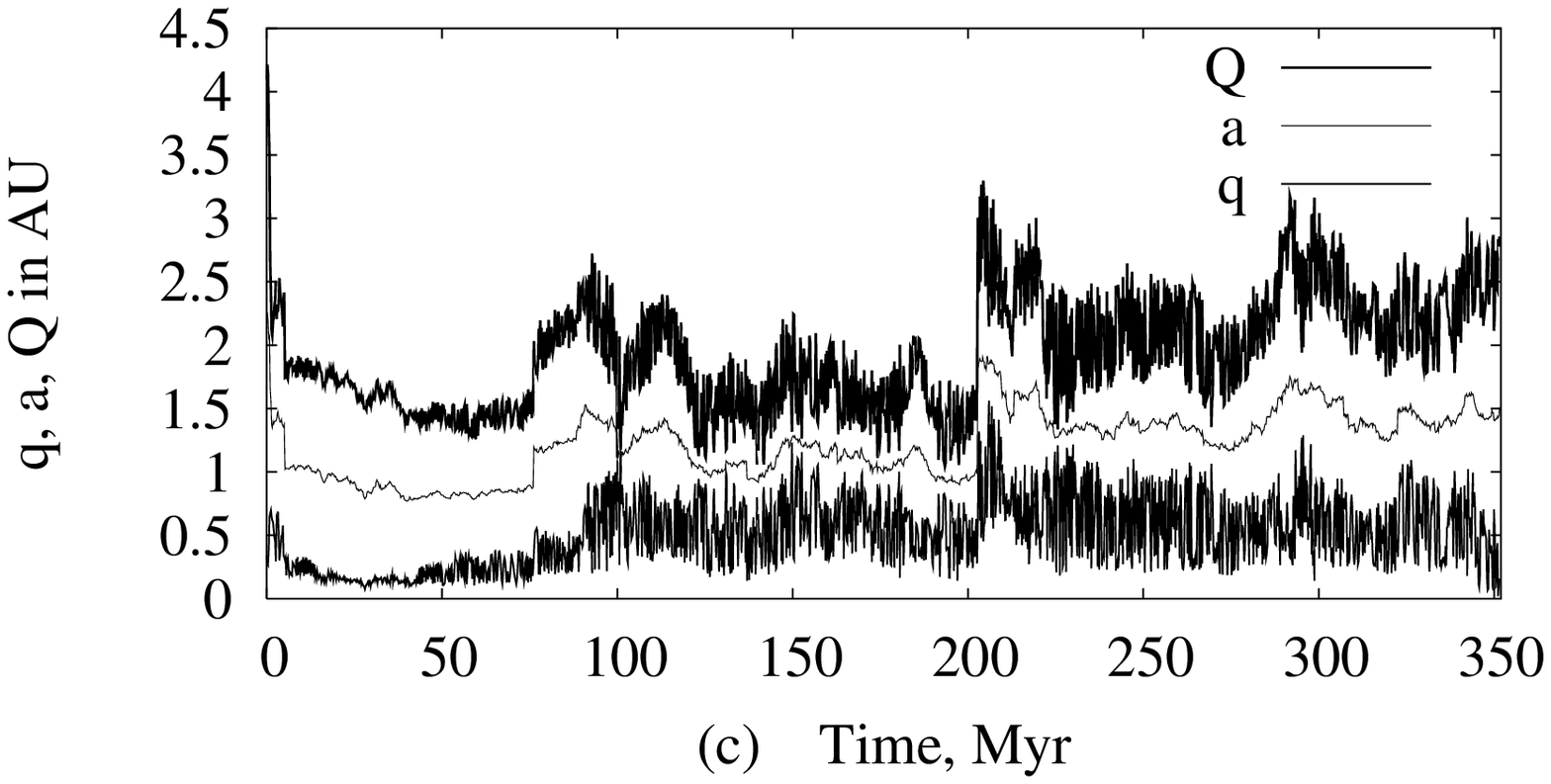}
\includegraphics[width=91mm]{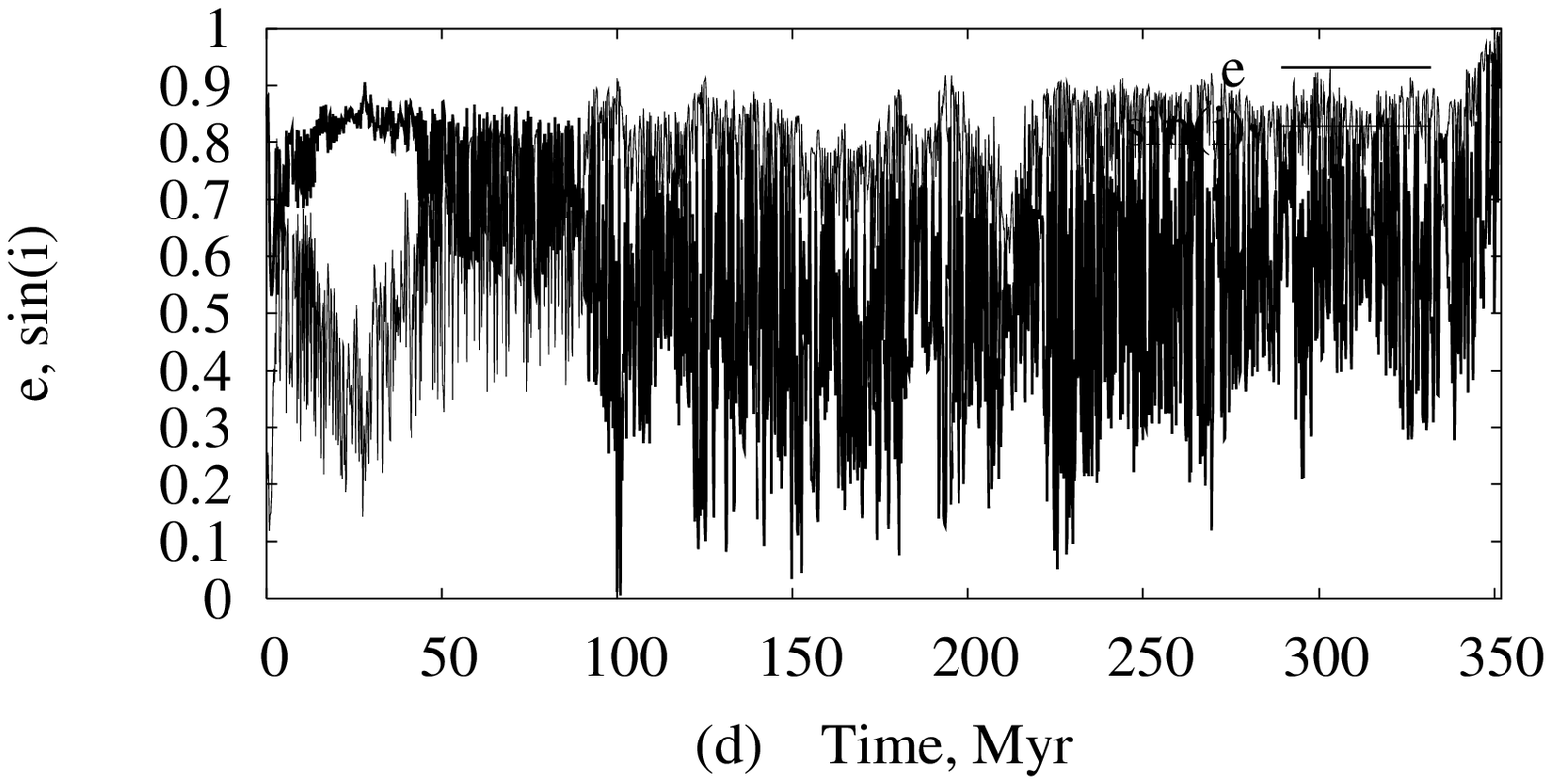}
\caption{Time variations in $a$, $e$, $q$, $Q$, sin($I$) for a former JCO in initial orbit close to that of Comet 10P (a-b) or Comet 2P (c-d). For (a) at $t<0.123$ Myr $Q>a>1.5$ AU.}
\end{figure}%

     In Fig. 2 we present the time in Myr during which objects had semi-major axes in the interval with a width of 0.005 AU (Figs. 2a-b) or 0.1 AU (Figs. 2c-d). Note that at 3.3 AU (the 2:1 resonance with Jupiter) there is a gap for asteroids that migrated from the 5:2 resonance and for former JCOs (except 2P).

\begin{figure}
\includegraphics[width=91mm]{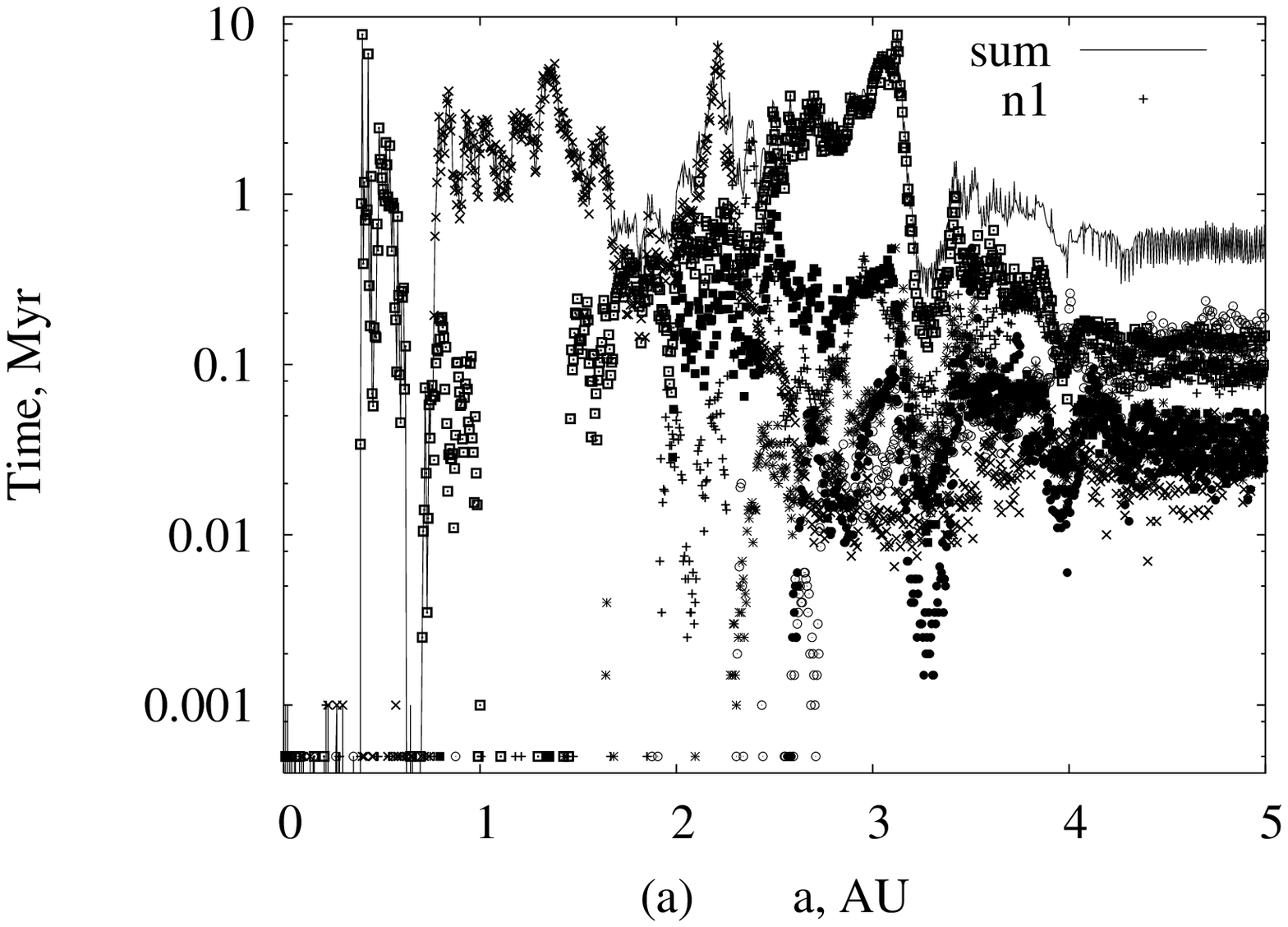}
\includegraphics[width=91mm]{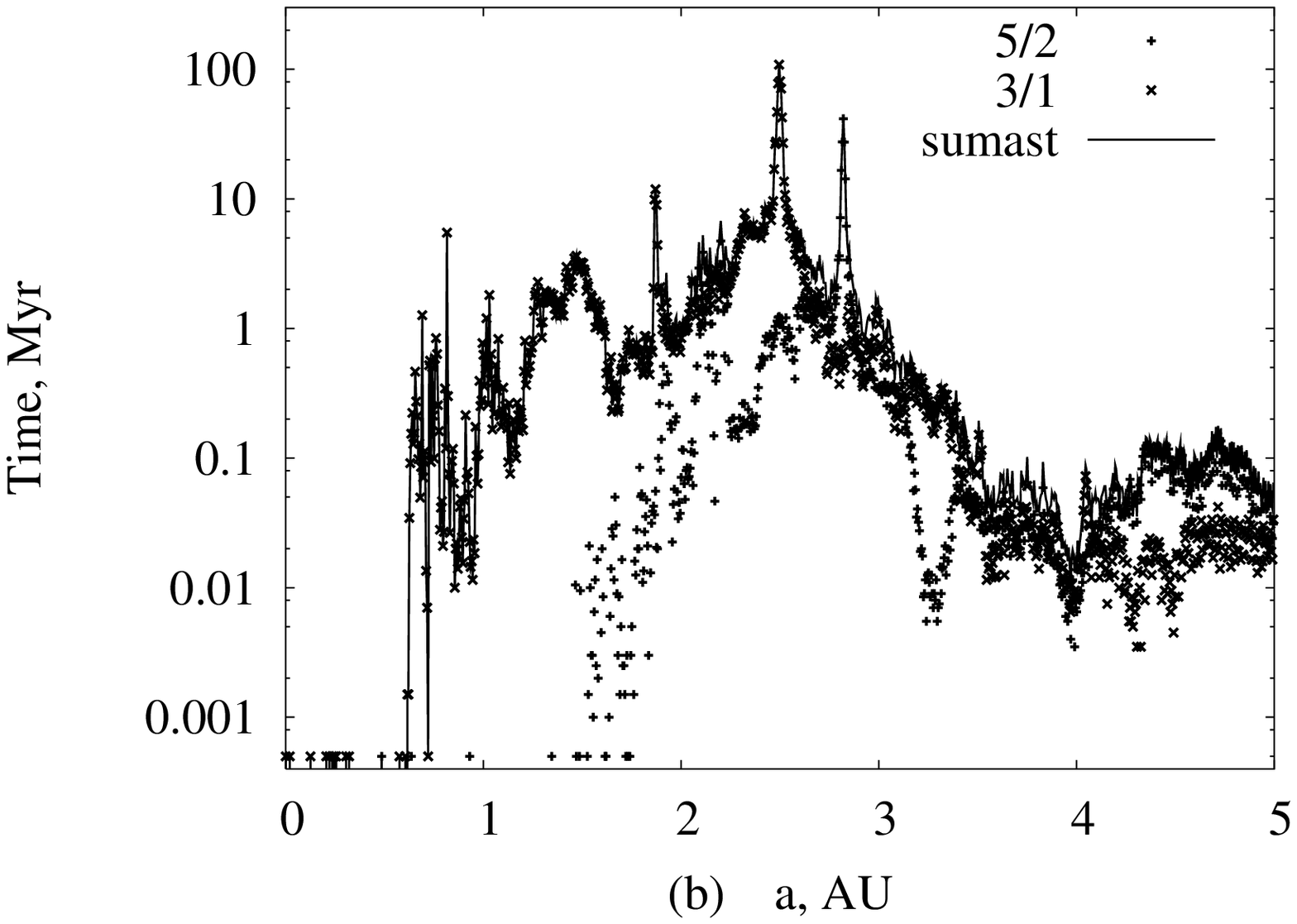} 
\includegraphics[width=91mm]{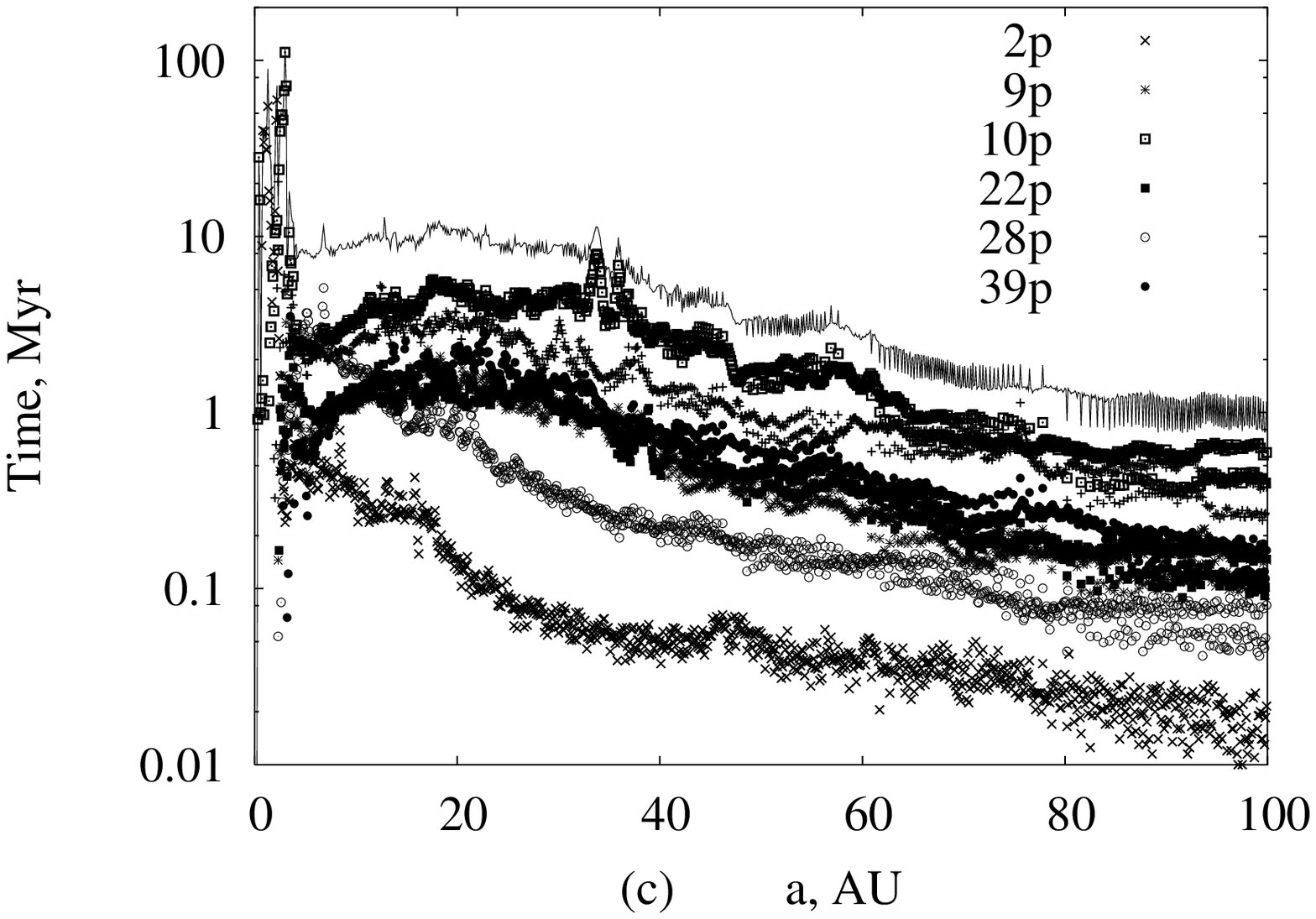}
\includegraphics[width=91mm]{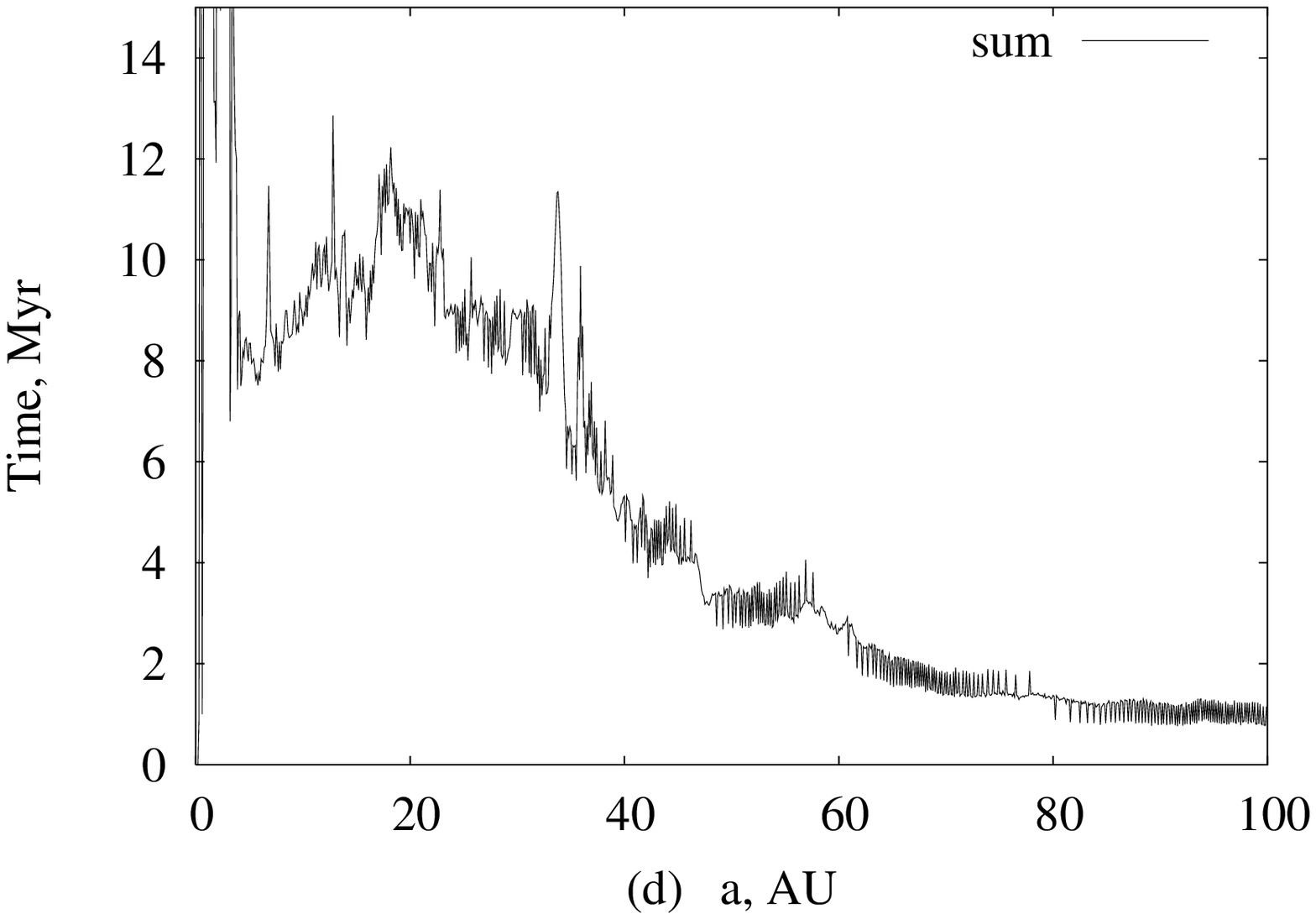}
\caption{ Distribution of migrating objects with their semi-major axes.
The curves plotted in (b) at {\it a}=40 AU  are (top-to-bottom) for sum,
10P, n1, 39P, 22P, 9P, 28P, and 2P. For Figs. (a) and (c), designations are the
same. } 
\end{figure}%

     The obtained probabilities of collisions of objects with planets show that the total mass of the matter delivered by short-period comets to an inner planet and normalized to its mass, for Mars and Venus can be greater than that for Earth. It would result in the relatively large ancient oceans on Mars and Venus.

     For $n1$, $T_J=0.12$ Myr and, while moving in Jupiter-crossing orbits, objects had orbital periods $P_a<10$, $10<P_a<20$, $20<P_a<50$, $50<P_a<200$ yr during about 11\%, 21\%, 21\%, and 17\% of $T_J$, respectively. So, the number of all JCOs is greater by a factor of 3 than that of Jupiter-family comets (which have period $<20$ yr). 

     Some JCOs, after residing in orbits with aphelia deep inside Jupiter's orbit, transfer for tens of Myr to the trans-Neptunian region, either in low or high eccentricitHy orbits. So some of the main asteroid belt bodies may get orbits of typical TNOs, then becoming SDOs having high eccentricities, and vice versa.

     The portion of bodies from the 5:2 resonance that collided with the Earth was by a factor of 7.6 smaller than that for the 3:1 resonance. Only a small portion of asteroids from the 5:2 resonance got $a<2$ AU (see Fig. 2b). 

\section*{3. COMPARISON OF THE RESULTS OBTAINED BY SYMPLECTIC AND DIRECT INTEGRATORS} 

     The results presented in Table 1, Figs. 1-2 and discussed above were obtained by the BULSTO code. We have also considered the orbital evolution of thousands of JCOs with the use of a symplectic method with an integration step $d_s$ equal to 3, 10, or 30 days. At $n1$ for 1997 JCOs (excluding 3 objects with maximum probability $P$ at $d_s=30$ days) the values of $P_r$ and $T$ are 6.0 and 8.2 (Venus), 4.6 and 11.9 (Earth), 0.83 and 18.9 (Mars), and $r$=2.5. These values are similar for Earth, larger for Venus, and smaller for Mars than those presenting in the first line of Table 1. About 1 among 1000 objects considered by RMVS3 at $d_s$=30 days got Earth-crossing orbits with $a<2$ AU during several tens of Myr (and even IEOs' orbits). These a few bodies increased the mean value of $P$ by a factor more than 10. At $d_s$=30 days four considered objects from the runs $n1$, 9P, 10P had a probability of collisions with the terrestrial planets more than 1 each (for 2P there were 21 such objects among 251 considered). Probably, the results of such symplectic runs can be considered as such migration that includes some nongravitational forces. For RMVS3, former JCOs got resonant orbits less often than for BULSTO and more often collided with the Sun. For example, for Comet 2P $P_S$=0.99 at $d_s$=10 days instead of 0.35 for BULSTO. 

     For resonant asteroids, we also obtained much larger values of $P$ and $T$ for RMVS3 at $d_s$=30 days than those for BULSTO. For asteroids initially located at the 3:1 resonance with Jupiter, $P_S$ was much larger for RMVS3 than for BULSTO:


$ \begin{array}{lllll} 

 & \varepsilon=10^{-9} & \varepsilon=10^{-8} & d_s=10 $ days$ & d_s=10 $ days$ \\

3:1 & 0.156 & 0.112 & 0.741 & 0.50 \\

5:2 & 0.062 & 0.028 & 0.099 & 0.155

\end{array} $


So in some cases a symplectic method gives a large error. We consider that for resonant asteroids and JCOs (especially in the cases when bodies can get close to the Sun) it is better not to use a symplectic method at all. Even if some results can be close to those by BULSTO, you don't know whether RMVS3 results are good enough before you compare the results with those obtained by direct integrations. Last years many scientists used a symplectic method for investigations of the orbital evolution of NEOs.

\section*{4. MIGRATION OF BODIES TO THE NEAR-EARTH SPACE}

According to [8], the fraction $P_{TNJ}$ of TNOs reaching Jupiter's orbit under the influence of the giant planets during 1 Gyr is 0.008-0.017. As mutual gravitational influence of TNOs can play a larger role in variations of their orbital elements than collisions [2], we consider the upper value of $P_{TNJ}$. Proceeding from the total of $5\cdot10^9$ 1-km TNOs within $30<a<50$ AU and assuming the mean time 0.13 Myr for a body to move in a Jupiter-crossing orbit, we obtain that about $10^4$ of former 1-km TNOs are now Jupiter-crossers and $3\cdot10^3$ of them are Jupiter-family comets. Using the total times spent by 7852 JCOs in various orbits (see Section 2), we obtain the following numbers of former 1-km TNOs now moving in several types of orbits:


$ \begin{array}{lllll} 

$IEOs$ & $Aten$ & $Al2$ & $Apollo$ & $Amor$ \\
100 &  860 & 4100 & 6600 & 1700

\end{array} $


As we considered mainly runs with relatively high migration to the Earth, the actual above values are smaller by a factor of several, the actual portion of IEOs and Aten objects can be smaller and that for Amor can be larger. Even if the number of Apollo objects is smaller by a factor of 10 than the above value, it is close to the real number (750) of 1-km Earth-crossing objects (half of them are in orbits with $a<2$ AU), although the latter number does not include those in high eccentric orbits. Eccentricities and inclinations of NEOs in our runs were larger than those for real NEOs, and probably most of former 1-km TNOs (extinct comets) now moving in NEOs' orbits with high eccentricities and inclinations have not yet been discovered, as most of the time they move relatively far from the Earth. For $P=6\cdot10^{-6}$, we obtain that former 1-km TNOs collide with the Earth once in 0.5 Myr. Note that the characteristic time elapsed up to a collision of an object with the Earth is $T_c=T/P\approx2.7$ Gyr for 7850 considered objects (1.1 Gyr for 7852 objects), while for observed Earth-crossing objects it is much less ($T_c\approx100$ Myr).

     The above estimates of the number of NEOs are approximate. For example, it is possible that the number of 1-km TNOs is less by a factor of several than $5\cdot10^9$, though some scientists considered that this number can be up to $10^{11}$. Also, the portion of TNOs migrated to the Earth can be smaller. On the other hand, the above number of TNOs was estimated for $a<50$ AU, and TNOs from more distant regions can also migrate inward. The Oort cloud also could supply Jupiter-family comets. According to [9], the rate of an object decoupling from the Jupiter vicinity and transfer to the NEO-like orbit is increased by a factor of 4 or 5 due to nongravitational effects. This would result in the larger values of $P_r$ and $T$ compared to those shown in Table 1. We consider that collisions of JCOs with asteroids could increase the number of objects migrating to the near-Earth space, especially to orbits with $a<2$ AU. The less are the masses of objects, the more often they can change their orbits due to collisions with smaller bodies. Our estimates show that, in principle, the trans-Neptunian belt can provide a considerable portion of Earth-crossing objects, but, of course, some NEOs came from the main asteroid belt. It may be possible to explore former TNOs near the Earth's orbit without sending spacecraft to the trans-Neptunian region.

     More than a half of the close encounters with the Earth belong to long-period comets, which amounts to about 80\% of the all known population. Thus, though probabilities are smaller for larger eccentricities, the number of collisions of both long-period and short-period active comets with the inner planets can be of the same order of magnitude, but most of former Jupiter-family comets can appear as typical asteroids and collide with the Earth from typical NEOs' orbits. 

     Based on the estimated collision probability $P=6\cdot10^{-6}$ (this value is a little larger than that for $n1$, but is smaller by a factor of 14 than that for a total 7852 JCOs) and assuming the total mass of planetesimals that ever crossed Jupiter's orbit is $\sim 100m_\oplus$ ($m_\oplus$ is mass of the Earth), we found that the total mass of bodies impacted on the Earth is $6\cdot10^{-4} m_\oplus$. If ices composed only a half of this mass, then the total mass of ices that were delivered to the Earth from the feeding zone of the giant planets turns out to be a factor of 1.5 greater than the mass of the Earth oceans ($\sim2\cdot10^{-4} m_\oplus$).

\section*{NOTES ABOUT THE PRESENT TEXT}

The text presented to the Proceedings was in Word and was printed in two columns.
After submitting the paper to the Proceedings, the orbital evolution of one JCO (presented in Figs. 1c-d) was calculated for a larger time interval.
For the astro-ph paper, I made a few corrections in Figs 1c-d, 2a,c-d and in text, which are connected with the new data for this JCO. 

\section*{ACKNOWLEDGEMENTS}

     This work was supported by INTAS (00-240), NRC (0158730), NASA (NAG5-10776), and RFBR (01-02-17540). The author is thankful to V.I. Ipatova for the help in calculations.

\end{document}